\documentclass[12pt]{arXiv_class}
\usepackage{authblk}

\newcommand\blfootnote[1]{%
  \begingroup
  \renewcommand\thefootnote{}\footnote{#1}%
  \addtocounter{footnote}{-1}%
  \endgroup
}

\newtheorem{remark}{Remark}

\title{\textbf{Full Optical Fiber Link Characterization \\ with the BSS-Lasso}}

\author{Raphael~Saavedra,~Pedro~Tovar,~Gustavo~C.~Amaral,~and~Bruno~Fanzeres}

\begin{document}
    \maketitle
	
	% ABSTRACT ============================================================================================
    
	\begin{abstract}

	In this work, a novel technique for optical fiber monitoring that introduces the Lasso as a signal processing technique within the Baseband Subcarrier Sweep (BSS) framework, called the BSS-Lasso, is proposed. The methodology is tested in simulated and real-world environments, taking into account both reflective and non-reflective events. The results show that, for fiber links ranging from 2 to 15 km with up to 3 faults, over 80\% of faults are detected within a 50 m range, and indicate that the proposed methodology significantly outperforms current state-of-the-art BSS-based supervision techniques. Finally, the BSS-Lasso allows for precise, low-cost, transmitter-embedded full characterization of optical fiber links.
	\end{abstract}
	%
	% INTRODUCTION =========================================================================================
	
	\blfootnote{
	
	R.~Saavedra is with the Department of Electrical Engineering and the Laboratory of Applied Mathematical Programing and Statistics (LAMPS), Pontifical Catholic University of Rio de Janeiro (PUC-Rio), Rio de Janeiro, Brazil (e-mail: rsaavedra@ele.puc-rio.br).

    P.~Tovar is with the Center for Telecommunications Studies, PUC-Rio (e-mail: ptovar@opto.cetuc.puc-rio.br).
    
    G.~C.~Amaral is with the Center for Telecommunications Studies, PUC-Rio and with QuTech and Kavli Institute of Nanoscience, Technical University of Delft, Delft, The Netherlands (e-mail: gustavo@opto.cetuc.puc-rio.br).
    
    B.~Fanzeres is with the Department of Industrial Engineering and LAMPS, PUC-Rio (e-mail: bruno.santos@puc-rio.br).
    
    Copyright (c) 2018 IEEE. Personal use of this material is permitted.  However, permission to use this material for any other purposes must be obtained from the IEEE by sending a request to pubs-permissions@ieee.org.
    }
	
	\section{Introduction}\label{Introduction}
	As society becomes more dependent on fast distribution of information, robust operation of telecommunication links appears as a top priority for network managers, since even the shortest service outage can affect up to millions of users; recently, it has been estimated that 80\% of all long-distance data traffic in the world is carried by optical fibers \cite{Kumar2014fiber}. The mechanical fragility of the fibers, however, poses a threat to the robust operation of the optical fiber links since fiber bending, fiber breaking, imperfect fiber splices, and corrupted connectors affect the power budget and, thus, transmission capacity. With the evolution of data transmission protocols and network architectures, the link supervision technology must also evolve to meet this sought-after robustness whilst maintaining a low impact over the quality of data transmission \cite{urban2013fiber}.
    
    The most successful and well-established supervision technique of the physical layer of optical networks is the Optical Time-Domain Reflectometry (OTDR) \cite{BarnoskiAO1977}. The main advantage of such technology is the possibility to obtain accurate precision with high resolution in long distance monitoring \cite{DericksonBOOK1998, liu2001events}. However, in order to fully acquire information from a fiber stretch using this technique, the transmission data is generally suspended for a non-negligible time frame since the OTDR pulse carries a considerable amount of optical power spread through a broad spectral bandwidth. As a consequence, although technically efficient, standard OTDR monitoring is usually economically burdening (see \cite{eraerds2010photon} and the references therein for a wider discussion).
    
	To tackle this issue, several other reflectometry-based techniques have been proposed in technical literature \cite{hangai1990detection, dong2015combined, AmaralJLT2015, amaral2017multiple, calliari2018high}. Generally, they seek for a monitoring routine such that a narrow spectral channel can be allocated, widening the simultaneous data traffic capability. In other words, an equilibrium between an efficient monitoring capacity, both in distance and resolution, and the coexistence with data transmission is pursuit. Thus, in practice, network operators need to balance the pros and cons of different monitoring techniques in order to choose the one that best fits their network, both from a technical and an economical perspective.
    
    Following this rationale, a novel transparent and cost-effective reflectometry-based technique has been recently developed \cite{amaral2017low}. Due to its characteristics, the technique will be henceforth dubbed as Baseband Subcarrier Sweep (BSS). From a technical point of view, the main goals of the BSS monitoring method are to achieve reasonable spatial resolution and dynamic range in optical fiber monitoring, with both negligible \textit{a priori} knowledge about the fiber and interference on data traffic. Within these four goals, three have been successfully achieved, with a dynamic range limited to 7 dB being the major hindrance of the method. Furthermore, from an implementation point of view, this technique fits the architecture of the so-called Mobile Fronthaul \cite{pizzinat2015things,urban2016fiber}, an ubiquitous concept for next-generation mobile networks, and can be seamlessly incorporated into the optical transmitter (a so-called transmitter-embedded technique) with minimum cost overhead \cite{urban2016fiber}.
    
    Generally speaking, the nature of the BSS-based monitoring technique is to measure the fiber's transfer function by detecting the Rayleigh backscattered portion of the propagating optical signal modulated by a swept tone covering several frequencies within an allocated low-frequency bandwidth. Even though this had been previously studied in \cite{nakayama1987optical}, it was only in \cite{amaral2017low} that the manipulation of the Rayleigh backscattered signal became the focus of the monitoring solution. The resulting probing signal has a periodic structure in the frequency domain, which can be represented as a linear combination of spatial-dependent phasors that take the fault positions as arguments \cite{amaral2017multiple}. Identifying which spatial-dependent phasors are present in the monitoring signal allows one to consequently identify the set of fault positions present in the fiber. Although intuitive, the key drawback of this methodology is the necessity to perform an extensive combinatorial search in order to precisely determine the fault positions, an utterly non-trivial task to be conducted by naive optimization methods \cite{amaral2017low, natarajan1995sparse} in reasonable computational time.
    
    In this context, techniques based on high-dimensional analysis emerge as an effective tool. In this work, a widely-used high-dimensional signal interpreter method known as Least Absolute Shrinkage and Selection Operator (Lasso) \cite{tibshirani1996regression} is employed in order to perform a computationally efficient fault detection \cite{parrilla1991digital}. Fundamentally, the Lasso performs a variable selection from an over-complete dictionary, electing the components that best fit the original signal. Therefore, by designing the over-complete dictionary with the appropriate sinusoidal-based functions associated to all possible fault positions (e.g., performing a meter-by-meter discretization of the fiber length), the Lasso methodology can be used to evaluate the optical fiber link in practical time. 
    
    In fact, the Lasso has already been adapted to fit in BSS-based monitoring techniques \cite{amaral2017multiple}. Nevertheless, the key issue regarding this previous work was the absence of a complete mathematical model describing the impact of reflective and non-reflective events on the acquired signal; only non-reflective events were considered within the dictionary. Characteristics of events allow their classification as reflective or non-reflective, and a correspondence with the underlying physical mechanism can be observed: fiber breaking and corrupted connectors usually causing reflective events; and fiber bending and imperfect fiber splices causing non-reflective events. Since the incidence of these cannot be predetermined, the absence of either causes the model to be incomplete. In order to compensate for the unaccounted contribution of reflective events, the incomplete model either produces unreal events or induces shifts on the real event positions, thus reducing the accuracy and robustness of the monitoring method. Therefore, the main objectives and contributions of this work are threefold:
    \begin{enumerate}
    	\item To extend the methodology proposed in \cite{amaral2017multiple} to account for a hybrid reflective and non-reflective modeling framework, which represents most practical cases of fiber link supervision. The proposed model makes use of the signal description derived in \cite{amaral2017low} to construct the over-complete dictionary that feeds the Lasso.
        %
%         \item To propose an extension of the Lasso methodology to accommodate an \textit{ex-post} analysis, with the goal of correcting imprecisions caused by reflective events. By making use of the properties of reflective events, a tailored heuristic, hereinafter referred to as \textit{BSS-Lasso}, is designed to enhance the fault position detection.
		\item To devise an \textit{ex-post} analysis to enhance the fault position estimation. By making use of the properties of reflective events, a tailored heuristic, hereinafter referred to as \textit{BSS-Lasso}, is proposed.
        \item To design and validate a methodology to properly reconstruct the time-domain profile of a fiber based on the BSS-Lasso method. The model that relates the frequency- to the time-domain profile allows precise estimation of the fault and reflection magnitudes, a feature that enables the full characterization of the fiber's profile.
    \end{enumerate}
    
   Finally, as a minor contribution, a large library of faulty fiber profiles in the frequency domain is made available so they can be used to perform extensive tests and comparisons with current state-of-the-art monitoring techniques. All data of this test bench is available in \cite{TestBenchIEEE}. This step, and the possibility of recreating frequency-domain profiles that mimic experimental acquired data is of great importance for future proposals and allows one to validate and compare the capacity of different monitoring routines.
    
    Due to the interdisciplinary character of this work, which involves optical backreflection measurements and high-dimensional data processing, an introductory background is provided. Firstly, the data acquisition and mathematical model of the fault location problem are briefly presented in Section \ref{sec:dataAcq_Model}, where the high-dimensional feature of the problem naturally arises from the characteristics of the supervision technique, and its constraints are pointed out. In Section \ref{sec:lasso}, the framework of the Lasso is introduced, along with a discussion on how the idiosyncrasies of the stated problem allow for the tailored heuristic, the BSS-Lasso, to be proposed. With the background from these two Sections, experimental fault location results using real-world fibers are presented in Section \ref{sec:exp_results}. Moreover, in Section \ref{sec:sim_results}, the comparison of the BSS-Lasso with state-of-the-art monitoring techniques is performed in the constructed test bench in order to provide a statistically relevant analysis of the technique's monitoring capability. The paper is concluded in Section \ref{sec:conclusion}, where the advances achieved by the BSS-Lasso and on-going research are summarized. % Introduction
    
    %------------------------------------------------

	\section{Experimental Setup} \label{sec:dataAcq_Model}    
    Sweeping the frequency of an optical sub-carrier tone to evaluate fault locations in an optical fiber has been extensively studied, and is referred to as Incoherent Optical Frequency-Domain Reflectometry (I-OFDR) \cite{Liehr2010_IOTDR}. In case the bandwidth of the frequency sweep is sufficiently large (in the order of GHz) to yield a granular spatial resolution (in the order of meters), a Fourier transform can be employed to extract the desired information \cite{pierce2000optical}, but at the high cost of consuming a significant portion of the available transmission band, except in specific cases where the transmission spectrum can be tailored to accommodate the monitoring signal \cite{urban2016fiber}. Conversely, if the bandwidth of the frequency sweep is reduced (e.g., to a few kilo Hertz), the achievable spatial resolution through a Fourier transform is not sufficient for a precise fault location, but allows for seamless adaptation into certain data transmission formats, with special stress to Sub-Carrier Multiplexed optical network architectures, where the baseband of the optical carrier is left unoccupied \cite{urban2016fiber, amaral2017low}. 
    
    The BSS technique falls within the latter class, making itself attractive from the point of view of adaptation to transmitters and coexistence with data transmission. The data acquisition is accomplished by modulating an optical carrier electric field with a sinusoidal signal, which has its frequency increased in a stepwise fashion that respects the steady-state condition of the fiber, while the amplitude and phase of the backscattered signal inside the frequency swept bandwidth are determined by a complex frequency beat detector.
    
    \subsection{Data Acquisition and Mathematical Model}
    
\begin{figure}[h]
		\centering
		\includegraphics[width=0.7\columnwidth]{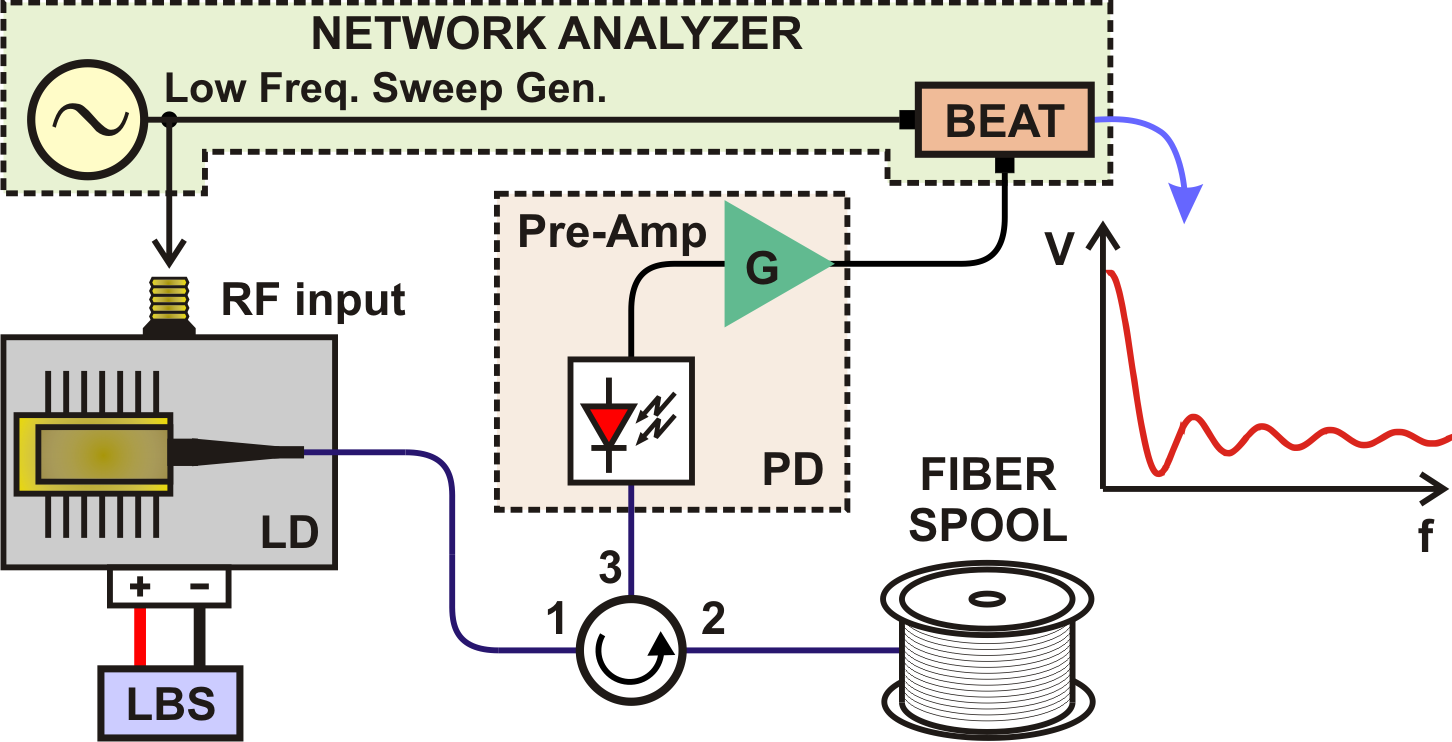}
		\caption{Experimental Setup of the baseband subcarrier sweep monitoring. Data acquisition is performed by measuring the steady-state amplitude and phase values for each  frequency step. LD: Laser Diode; NA: Network Analyzer; PD: Photodiode; LBS: Laser Bias Source; BEAT: complex frequency beat detector.}
		\label{fig:exp_setup}
	\end{figure}

    The experimental setup for data acquisition is presented in Fig. \ref{fig:exp_setup}. A Network Analyzer (NA) generates the low-frequency step-wise swept tone that directly modulates the current of a laser diode biased with a Laser Bias Source (LBS). This causes the electric field of the optical carrier to be modulated accordingly, thus creating a low-frequency optical sub-carrier channel with bandwidth defined by the step-wise sweep range, which propagates through the fiber; this channel is henceforth dubbed the \textit{monitoring channel}. The optical circulator, placed immediately before the Fiber Under Test (FUT), allows one to direct any counter-propagating signal, such as the Rayleigh backscattered portion of the incoming signal, to a photodetector. Due to the elastic character of Rayleigh scattering, the detected backscattered signal will conserve the original modulation, so that a frequency analysis inside the low-frequency optical sub-carrier channel can be performed with the correct apparatus. In this case, the NA itself allows for complex (amplitude and phase) frequency beat detection, so the resulting electrical signal is amplified and directed to the NA for frequency response analysis. 
    
    A key challenge to adapt BSS-based technologies in fiber monitoring is precisely the spatial resolution limitation imposed by the Fourier transform analysis induced by a frequency sweep with limited range. As discussed in \cite{amaral2017low}, analysis of the resulting signal directly in the frequency domain permits one to overcome such limitation and extend the achieved spatial resolution. In order to model the backscattered signal inside the monitoring channel bandwidth, the fact that the OTDR profile $\left\{P\left(z\right)\right\}_{z \in [0,L]}$ of a fiber with length $L$ meters can be suitably approximated by a linear combination of step (breaks) and peak (reflections) functions with a single slope (the fiber's attenuation coefficient) is used \cite{WeidJLT2016}. More specifically, $\forall ~ z \in [0,L]$,
    \begin{align}
    \begin{split}
		P\left(z\right) = e^{-2\alpha z} &\left(\sum_{b \in \mathcal{B}} \phi_{b} \Big[ u\left( z \right) - u\left( z - X_{b} \right) \Big] \right. + \left.\sum_{r \in \mathcal{R}} \theta_{r} \Big[ \delta\left(z - X_{r} \right) \Big]\right), \label{eq:PzHeaviside}
    \end{split}
	\end{align}
    where $\left\{ X_{b} \right\}_{b \in \mathcal{B}}$ are the non-reflective events with $\mathcal{B}$ the set of non-reflective events indexes, and $\phi_b$ are the linear combination coefficients. Similarly, $\left\{X_{r}\right\}_{r \in \mathcal{R}}$ are the reflective events with $\mathcal{R} \subseteq \mathcal{B}$ the set of reflective events indexes, and $\theta_r$ are the linear combination coefficients. In \eqref{eq:PzHeaviside}, $u\left(z\right)$ and $\delta\left(z\right)$ denote, respectively, the Heaviside and Dirac impulse functions.

    Therefore, the fiber profile contains the level shifts that correspond to non-reflective faults and spikes that correspond to reflective events. Note that, although $P\left(z\right)$ translates the amount of power being backscattered and/or reflected from each position of the fiber, the coefficients $\phi_b$ are not directly related to the faults magnitudes as usually presented in an OTDR; this relation is given by
    \begin{equation}
    	\xi_{b} = \sqrt{1 - \frac{\phi_b}{\displaystyle \prod_{j \in \mathcal{B} | j < b} \xi_{j}^{2}}}, \hspace{1.00cm} \forall ~ b \in \mathcal{B},
    	\label{eq:magCoeff}
    \end{equation}
    where $\xi_b$ are the fault magnitudes of the OTDR profile in linear scale. It is clear from Eq. \eqref{eq:magCoeff} that it is possible to recursively calculate the fault magnitudes $\{\xi_b\}_{b \in \mathcal{B}}$ given $\{\phi_b\}_{b \in \mathcal{B}}$ \cite{amaral2017low}.
    
    As the modulated electric field of the optical carrier propagates into the fiber, the backscattered signal (conserving the modulation properties) will propagate in the opposite direction. Since the modulating function is a sinusoidal wave, the modulated optical electric field can be cast into a phasor form with parameter $\kappa$, where $\kappa$ is the wave vector $\kappa = \tfrac{\omega n}{c}$, with $c$ denoting the speed of light, $n$ the group index of refraction of the fiber's core, and $\omega$ the low-frequency modulation of the electric field. Due to the fact that the signal is analyzed after the opto-electrical conversion, the optical phasor with optical wave-vector will be omitted \cite{amaral2017low}. The electrical power at a specific frequency $S\left( f \right)$, detected after the optical circulator, will correspond to the integral of $P\left(z\right)$ over $z \in [0, L]$ multiplied by the modulating function. Formally,
    \begin{align}
    \begin{split}
		S\left(f\right) &= \int_{z \in [0, L]} P(z) A e^{j2 \kappa z}dz \\
        				&= \int_{z \in [0, L]} \left(\sum_{b \in \mathcal{B}} \phi_{b} \Big[ u\left(z\right) - u\left( z - X_{b} \right) \Big] \right. + \\
		& \hspace{1.6cm} \left. \sum_{r \in \mathcal{R}} \theta_{r} \Big[ \delta\left( z - X_{r} \right) \Big] \right) A e^{j2z(\kappa + j\alpha)}dz, 
    \label{eq:Sf_IntForm}
    \end{split}
	\end{align}
     where it should be noted that, for convenience, the intrinsic attenuation coefficient of the optical fiber has been manipulated to appear in the exponential term.
     
     For presentation purposes, experimental parameters on which the measured backscattered signal depends, such as the Rayleigh backscattering coefficient of the fiber $C$, the photodiode's responsivity $R$, the power launched into the fiber $P_{0}$, the portion of the modulation depth occupied by the monitoring signal $m$, the photodetector gain $G$, and the Noise-Equivalent Power (NEP) of the Network Analyzer $N_{na}$, are merged into $A = C R P_0 m G N_{na} $, for simplicity, as they are assumed to be constant throughout the measurements. The role of each of these parameters will be discussed with depth in Section \ref{sec:exp_results}.
    
    The integral in equation \eqref{eq:Sf_IntForm} can be solved analytically by making use of properties of the Heaviside and Dirac impulse functions. More precisely, the Heaviside steps change only the limits of the integral, while the Dirac impulses evaluate the multiplying function at its displacement point. The result is the following complex function, that enables one to associate the frequency profile characteristics of the fiber to both reflective and non-reflective events.
    \begin{align}
		S\left(f\right) &= \sum_{b \in \mathcal{B}} \Phi_{b} S_{B} \left( f, X_{b} \right) + \sum_{r \in \mathcal{R}} \Theta_{r} S_{R}\left( f, X_{r} \right), \label{eq:SfFinal}
    \end{align}
    where
    \begin{align}
    \begin{split}
    	\Phi_{b} = A \phi_{b}; &\hspace{1.00cm} \Theta_{r} = A \theta_{r};\\[0.5em]
    	S_{B} \left( f, X_{b} \right) &= \left(\frac{e^{j \frac{4 \pi f n}{c} X_{b}} e^{-2 \alpha X_{b}} - 1}{j \frac{4 \pi f n}{c} - 2 \alpha}\right) ; \\[0.5em]
        S_{R}\left( f, X_{r} \right) &= \left( e^{j \frac{4 \pi f n}{c} X_{r}} e^{-2 \alpha X_{r}} \right). \label{eq:NonReflectiveTerm}
     \end{split}
     \end{align}
     
     The non-reflective $\left( S_{B} \right)$ and reflective $\left(S_{R}\right)$ terms in \eqref{eq:NonReflectiveTerm} correspond to the so-called spatial-dependent phasor \cite{amaral2017low, amaral2017multiple}. The representation in \eqref{eq:SfFinal} is more suitable for monitoring applications, since one can explore the fiber length $[0,L]$ to construct the sets $\mathcal{B}$ and $\mathcal{R}$ that adequately recover the acquired signal, thus precisely indicating the fault positions. In Section \ref{sec:lasso}, an efficient methodology to identify the sets $\mathcal{B}$ and $\mathcal{R}$ on \eqref{eq:SfFinal} based on the Lasso is presented.
       
    \subsection{Equivalence of time- and frequency-domain monitoring} \label{Sec:Eq_TimeFreq}
    
    Following the results of the previous section, it becomes clear that identifying the sets $\mathcal{B}$ and $\mathcal{R}$, which corresponds to fully characterizing the optical fiber, can be performed either in the frequency domain using $S\pars{f}$ and the low-frequency profile \cite{amaral2017multiple} or in the time domain using $P\pars{z}$ \cite{WeidJLT2016}. Developing a consistent mathematical procedure that identifies these sets while utilizing the low-frequency profile of the fiber is the main goal of this work. Before the method is described, however, it is interesting to analyze the mathematical model and validate its description. For this purpose, the OTDR is an excellent reference since it is the standard procedure for characterizing a fiber link, as it provides the time-domain profile of the fiber.
    
    Given a fiber link for which the reflective and non-reflective events have been identified, the validation of the mathematical model can be performed by evaluating the low-frequency response of the fiber following two different -- but, in principle, equivalent -- paths: (i) the signal output from the Network Analyzer, i.e., the low-frequency profile of the fiber, can be compared to the model function $S\pars{f}$ using the known sets $\mathcal{B}$ and $\mathcal{R}$; and (ii) the OTDR profile of the fiber $\left(\{P\pars{z}\}_{z \in [0,L]}\right)$ combined with Eq. \eqref{eq:Sf_IntForm} generates the low-frequency profile, which can also be compared to the model function $S\pars{f}$ given $\mathcal{B}$ and $\mathcal{R}$. The agreement between these two branches of analysis validates the model and allows for a direct translation between the results in frequency and time domains.
    
\begin{figure}[h]
		\centering
		\includegraphics[width=0.7\columnwidth]{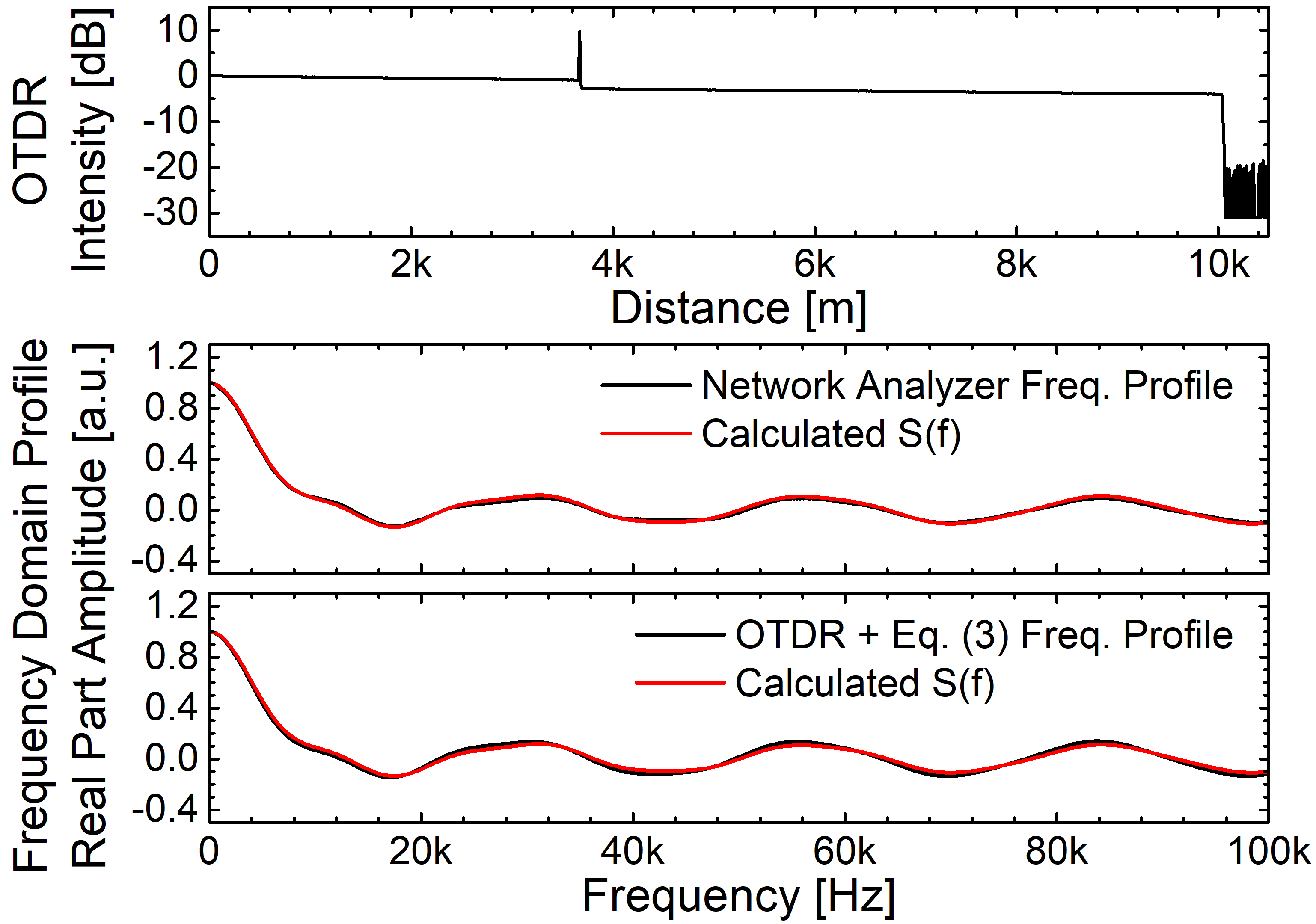}
		\caption{Time- and frequency-domain reconciliation based on the two aforementioned paths of analysis. In the upper panel, the time-domain profile is depicted as a reference. In the second panel, the output of the BSS monitoring technique (black) is compared to the model function (red). In the third panel, the frequency-domain profile calculated from the time-domain profile and \eqref{eq:Sf_IntForm} (black) is compared to the model function (red).}
		\label{fig:modValid}
	\end{figure}

    In the first panel of Fig. \ref{fig:modValid}, the OTDR profile, or time-domain profile, of an illustrative fiber is depicted as a reference; this trace is visually useful since the reflective and non-reflective events are directly identifiable. The respective positions are determined from it to be 3664 m and 10036 m. The frequency-domain profiles of the same fiber are depicted: in the second panel of Fig. \ref{fig:modValid}, following path (i); in the third panel of Fig. \ref{fig:modValid}, following path (ii). Together with the experimentally acquired frequency-domain profiles of the example fiber (in black), the model function $S\pars{f}$, calculated based on the known events positions, is also depicted (in red). It should be noted that only the real part of the profiles is depicted for ease of visualization even though its imaginary counterpart follows a similar pattern. Apart from slight mismatches arising from the imperfection of the measurement apparatus, the correspondence between the results validates the mathematical model described in \eqref{eq:Sf_IntForm}--\eqref{eq:SfFinal}.
     % Data Acquisition and Mathematical Model
    
    %------------------------------------------------
    
    \section{BSS-Lasso monitoring technique} \label{sec:lasso}
    The main objective of this section is to develop an efficient methodology to identify the spatial-dependent phasors that compose the frequency-domain profile acquired using the BSS probing apparatus. More specifically, the methodology aims to determine the set of non-reflective and reflective event locations, respectively $\mathcal{B}$ and $\mathcal{R}$, based on the data acquisition architecture described in Fig. \ref{fig:exp_setup}. For this purpose, the theoretical results developed in Section \ref{sec:dataAcq_Model} are combined with a high-dimensional technique based on $\ell_1$ regularization, known as Lasso \cite{tibshirani1996regression}. The proposed BSS-Lasso method involves three main stages thoroughly described in this section. The first stage, called \textit{selection stage}, extends the model constructed in \cite{amaral2017multiple} to account for reflective events; the second, named \textit{correction stage}, exploits properties of the problem to design an heuristic that corrects errors related to reflective events; finally, the third stage, called \textit{treatment stage}, performs a refinement step in order to output unambiguous results. For completeness purposes, we begin by presenting the foundations of the Lasso.
    
    \subsection{The Lasso technique} \label{sec:StandardLasso}
    
    The Lasso \cite{tibshirani1996regression} was originally designed to perform a computationally efficient variable selection in regression analysis through $\ell_1$ regularization (see \cite{Hastie2009_ElemStatLearn, Peres2018_Lasso_LitReview} for a systematic literature review on variable selection techniques and extensions of the Lasso). The goal of the Lasso is, thus, to provide an adequate selection of predictors in order to successfully explain the dynamics of a given variable while reducing the number of selected explanatory variables through the $\ell_1$-norm penalty. Formally, it consists of a standard least-squares formulation with the addition of a penalty factor on the $\ell_1$-norm of the selections:
    \begin{align}
		\min_{\boldsymbol{\beta}} \left\{ \big\|\mathbf{y} - \mathbf{M} \boldsymbol{\beta} \big\|^{2}_{2} + \lambda \> \big(\boldsymbol{\beta}^{\top} \mathbf{w}\big) ~ \middle| ~ \boldsymbol{\beta} \geq \boldsymbol{0} \right\}, \label{eq:lasso}
	\end{align}
    where $\mathbf{y}$ is the observed/dependent variable, $\mathbf{M}$ stands for a matrix of explanatory/independent variables, $\boldsymbol{\beta}$ is a decision vector of coefficients, and $\mathbf{w}$ and $\lambda$ are positive vector- and scalar-size parameters, respectively. 
    
    The $\ell_1$-norm penalty term (second term of the objective function) performs a regularization of the $\boldsymbol{\beta}$ vector around zero. In other words, it penalizes ``unnecessary'' deviations from zero, acting, thus, as a variable selection. In this regularization, $\mathbf{w}$ stands for the penalty weight at each coordinate of $\boldsymbol{\beta}$ and $\lambda$ defines the penalty magnitude for the total regularization term. It is important to note that the standard Lasso does not restrict the vector $\boldsymbol{\beta}$ to be positive. Nevertheless, since a permanent power loss or a momentary power peak (induced by a fault or reflection, respectively) can only generate complex frequency-domain signals that, under cartesian representation, exhibit positive coefficients, the positiveness constraint ensures the physical soundness of the model.
    	
	A key challenge to apply the Lasso in practical applications is an adequate definition of the penalty weights $\mathbf{w}$ and $\lambda$. Usually, practitioners set $\mathbf{w} = \mathbbm{1}$ and apply an information criterion (e.g., the Extended Bayesian Information Criterion (EBIC) \cite{smith1980bayes,chen2012extended}) to select, from a pre-defined set $\Lambda$ of penalty values, the magnitude of $\lambda$. Therefore, one needs to solve the optimization problem \eqref{eq:lasso} for all $\lambda \in \Lambda$ and pick the one that results in the best (i.e., minimal) EBIC. Algorithm \ref{alg:lasso} showcases the Lasso procedure as conducted in this work\footnote{The Lasso can be efficiently executed through several open-source packages, among which we highlight \texttt{glmnet} \cite{friedman2009glmnet}, that utilizes coordinate descent with Fortran subroutines and is available in several languages such as Julia, Matlab and R.}.
    \begin{algorithm}[H]
		\caption{Lasso}
		\begin{algorithmic}
			\label{alg:lasso}
			\FOR {$\lambda$ in $\Lambda$}
			\STATE $\hat{\boldsymbol{\beta}}(\lambda) \gets \arg \min_{\boldsymbol{\beta} \geq 0} \|\mathbf{y} - \mathbf{M} \boldsymbol{\beta} \|^{2}_{2} + \lambda \big(\boldsymbol{\beta}^{\top} \mathbf{w} \big) $
			\ENDFOR
			\RETURN $\hat{\boldsymbol{\beta}} \gets \arg \min_{\hat{\boldsymbol{\beta}}(\lambda)} \text{EBIC}(\hat{\boldsymbol{\beta}}(\lambda)) $
		\end{algorithmic}
	\end{algorithm}
    
    For the sake of simplicity, $\text{EBIC}(\hat{\boldsymbol{\beta}}(\lambda))$ represents the evaluation of the Extended BIC from the solution of \eqref{eq:lasso} with penalty $\lambda$. Henceforth, Algorithm \ref{alg:lasso} will be referred to as $\text{Lasso}(\mathbf{y},\mathbf{M},\mathbf{w})$.
    
	\subsection{Model design and selection stage of the BSS-Lasso} \label{sec:ModelDesign}
	
	In order to identify the set of fault locations on a given fiber link based on its frequency-domain profile, equation \eqref{eq:SfFinal} is accomodated into the Lasso framework discussed in Subsection \ref{sec:StandardLasso}. The procedure starts by creating a spatial discretization of the fiber length $L$ in $\mathcal{X} = \{X_{1}, \cdots, X_{q}\}$ locations. For a reasonable granularity, the fault locations are within a negligible distance of given elements in $\mathcal{X}$. Additionally, a byproduct of the data acquisition architecture discussed in Section \ref{sec:dataAcq_Model} is a set of frequencies $\mathcal{F} = \{f_{1}, \cdots, f_{m}\}$ for which the probing signal has been precisely evaluated. Therefore, the matrix $\mathbf{M}$ in \eqref{eq:lasso} can be defined as
	\begin{equation} \label{eq:M}
		\mathbf{M} = \frac{1}{L} \cdot \Big[~ \mathbf{M}_{B} ~ \Big| ~ \mathbf{M}_{R} ~\Big],
	\end{equation}
    where matrices $\mathbf{M}_{B}$ and $\mathbf{M}_{R}$ represent the fault and reflection signals as follows:
    \begin{align}
      \mathbf{M}_{B} = 
      \begin{bmatrix}
        \text{Re}\{S_{B}(\mathcal{F}, \mathcal{X})\}
        \\
        \vspace{-0.08in}
        \\
        \text{Im}\{S_{B}(\mathcal{F}, \mathcal{X})\}
      \end{bmatrix} \hspace{0.05cm};\hspace{0.05cm}
      \mathbf{M}_{R} = \begin{bmatrix}
        \text{Re}\{S_{R}(\mathcal{F}, \mathcal{X})\}
        \\
        \vspace{-0.08in}
        \\
        \text{Im}\{S_{R}(\mathcal{F}, \mathcal{X})\}
      \end{bmatrix}.
	\end{align}
	
    In \eqref{eq:M}, $\mathbf{M}$ contains the real and imaginary parts of the signals $S_{B}$ and $S_{R}$ defined in \eqref{eq:NonReflectiveTerm}, generated by every possible fault and reflection location within the given fiber discretization in $\mathcal{X}$ and frequency in $\mathcal{F}$. Furthermore, a normalization procedure (division by $L$ in \eqref{eq:M}) is performed to avoid numerical issues and an intercept with associated zero penalty can be added to account for possible measurement effects that offset the acquired data. According to the model description, the dependent variable $\mathbf{y}$ is an instance of $S(f)$, which is also decomposed into its real and imaginary parts:
    \begin{align}
     \mathbf{y} = 
      \begin{bmatrix} 
          \text{Re}\{S(\mathcal{F})\}\\
          \vspace{-0.08in}\\
          \text{Im}\{S(\mathcal{F})\}
      \end{bmatrix}. \label{eq:y}
    \end{align}
    
    Within this design, the decision vector $\boldsymbol{\beta}$ measures the coefficients $\Phi$ and $\Theta$ in equation \eqref{eq:SfFinal}, thus being decomposed into non-reflective and reflective coefficients: $\boldsymbol{\beta} = [ \{ \beta^{B}_{b} \}_{b = 1}^{q}, \{ \beta^{R}_{r} \}_{r = 1}^{q} ]^{\top}$. Therefore, the position within $\mathcal{X}$  in which a fault is located can be identified by picking $\{ \beta^{B}_{b} \}_{b = 1}^{q}$ greater than zero after executing Algorithm \ref{alg:selection_stage} with $\mathbf{M}$ and $\mathbf{y}$ defined in \eqref{eq:M} and \eqref{eq:y}, respectively. Note that, as a byproduct of the proposed methodology, the network operator can also distinguish if the fault position has a reflective event, i.e., if the corresponding $\beta^{R}$ is also greater than zero. The selection stage is explicitly presented in Algorithm \ref{alg:selection_stage}.
    \begin{algorithm}[H]
		\caption{BSS-Lasso -- Selection stage}
		\begin{algorithmic}
        \label{alg:selection_stage}
			\STATE $\hat{\boldsymbol{\beta}}^{(1)} \gets \text{Lasso}(\mathbf{y}, \mathbf{M}, \mathbbm{1})$
			\RETURN $\hat{\boldsymbol{\beta}}^{(1)}$
		\end{algorithmic}
	\end{algorithm}
    \begin{remark}
    	The monitoring technique discussed in \cite{amaral2017multiple}, called \textit{SincLasso}, is a particular instance of the selection stage by dropping matrix $\mathbf{M}_{R}$ from the model, thereby not taking reflective events into account.
    \end{remark}
    
	\subsection{Correction stage of the BSS-Lasso} \label{sec:BSS_Lasso}
	
    In \cite{amaral2017multiple}, it was observed that the proposed SincLasso monitoring scheme regularly failed to accurately identify fault positions whenever the event had a reflective component. More specifically, significant shifts in the estimated fault position with respect to the real position were observed. Furthermore, a similar pattern persisted in the BSS-Lasso selection stage, even with the explicit inclusion of reflective events in the modeling design. Nevertheless, it was observed that, although the fault selection remained shifted in the presence of a reflection, its reflective counterpart was consistently accurate. Therefore, by making use of this empirical observation, a correction stage was designed to accommodate an \textit{ex-post} analysis of the selection stage in order to improve the supervision of fibers with reflective events.
    
    The correction stage has a similar motivation to the Adaptive Lasso \cite{zou2006adaptive}, in which it utilizes Lasso selections to modify the penalty vector $\mathbf{w}$. More precisely, reflections only exist accompanied by a fault at the same position (i.e., $\mathcal{R} \subseteq \mathcal{B}$). Therefore, since the selection stage has shown much greater accuracy for the reflective selections, an adjustment on the weight vector $\mathbf{w}$ can be made at the respective fault position, reducing its penalty so that Algorithm \ref{alg:lasso} can be re-executed with the adjusted $\mathbf{w}$. The correction stage is presented next.
    
    \begin{algorithm}[H]
		\caption{BSS-Lasso -- Correction stage}
		\begin{algorithmic}
        \label{alg:correction_stage}
        	\STATE Let $\hat{\boldsymbol{\beta}}^{(1)}$ be the selection stage output.
        	\STATE Let $\mathcal{Q} = \{q+1, ..., 2q\}$, $\epsilon > 0$, $0 < \gamma < 1$.
			\STATE Initialize penalty vectors $\mathbf{w}^{(k)} \gets \mathbbm{1}$, $k = 2, 3$.
            \FOR{$k = 2$ \TO $3$}
            \STATE \textbf{i.} If $\max_{j \in \mathcal{Q}}\{\hat{\beta}_{j}^{(k-1)}\} = 0 \implies$ \textbf{return} $\hat{\boldsymbol{\beta}}^{(k-1)}$
			\STATE \textbf{ii.} Adjust penalty vector $\mathbf{w}^{(k)}$:
           	\STATE $w^{(k)}_{i - q} \gets \gamma, ~~ \forall ~ i \in \mathcal{Q} ~|~\hat{\beta}^{(k-1)}_{i} > \epsilon \cdot \max_{j \in \mathcal{Q}}\{\hat{\beta}_{j}^{(k-1)}\}$
            \STATE \textbf{iii.} Run the Lasso with penalty $\mathbf{w}^{(k)}$:
			\STATE $\hat{\boldsymbol{\beta}}^{(k)} \gets \text{Lasso}(\mathbf{y},\mathbf{M},\mathbf{w}^{(k)}) $
            \ENDFOR
			\RETURN $\hat{\boldsymbol{\beta}}^{(3)}$
		\end{algorithmic}
	\end{algorithm}
    
    In Algorithm \ref{alg:correction_stage}, $\epsilon$ is a small number that acts as a sensitivity threshold. Recall from the model design in Subsection \ref{sec:ModelDesign} that coordinates $\{1,...,q\}$ in $\boldsymbol{\beta}$ correspond to non-reflective event candidates $(\{\beta^{B}_{b} \}_{b = 1}^{q})$ and $\{q + 1,...,2q\}$ the reflective event candidates $(\{ \beta^{R}_{r} \}_{r = 1}^{q})$. Therefore, $\gamma \in (0,1)$ is defined to reduce the penalty weight on the first $q$ elements of $\boldsymbol{\beta}$ according to the reflective selections of the previous iteration, such that $\mathbf{w}^{(k)}$ has all elements equal to one, except those that represent faults with non-negligible reflective selections, which receive a reduced penalty value $\gamma$. More precisely, if the selection stage outputs any non-zero reflective selections, the BSS-Lasso enters the correction stage, where up to two more iterations are computed in order to address the previously mentioned shifting phenomenon. For each iteration $k$, a reduced penalty vector $\mathbf{w}^{(k)}$ is created based on the selections of the previous iteration, and the Lasso is executed accordingly.
    
\begin{figure}[h]
		\centering
		\includegraphics[width=0.7\columnwidth]{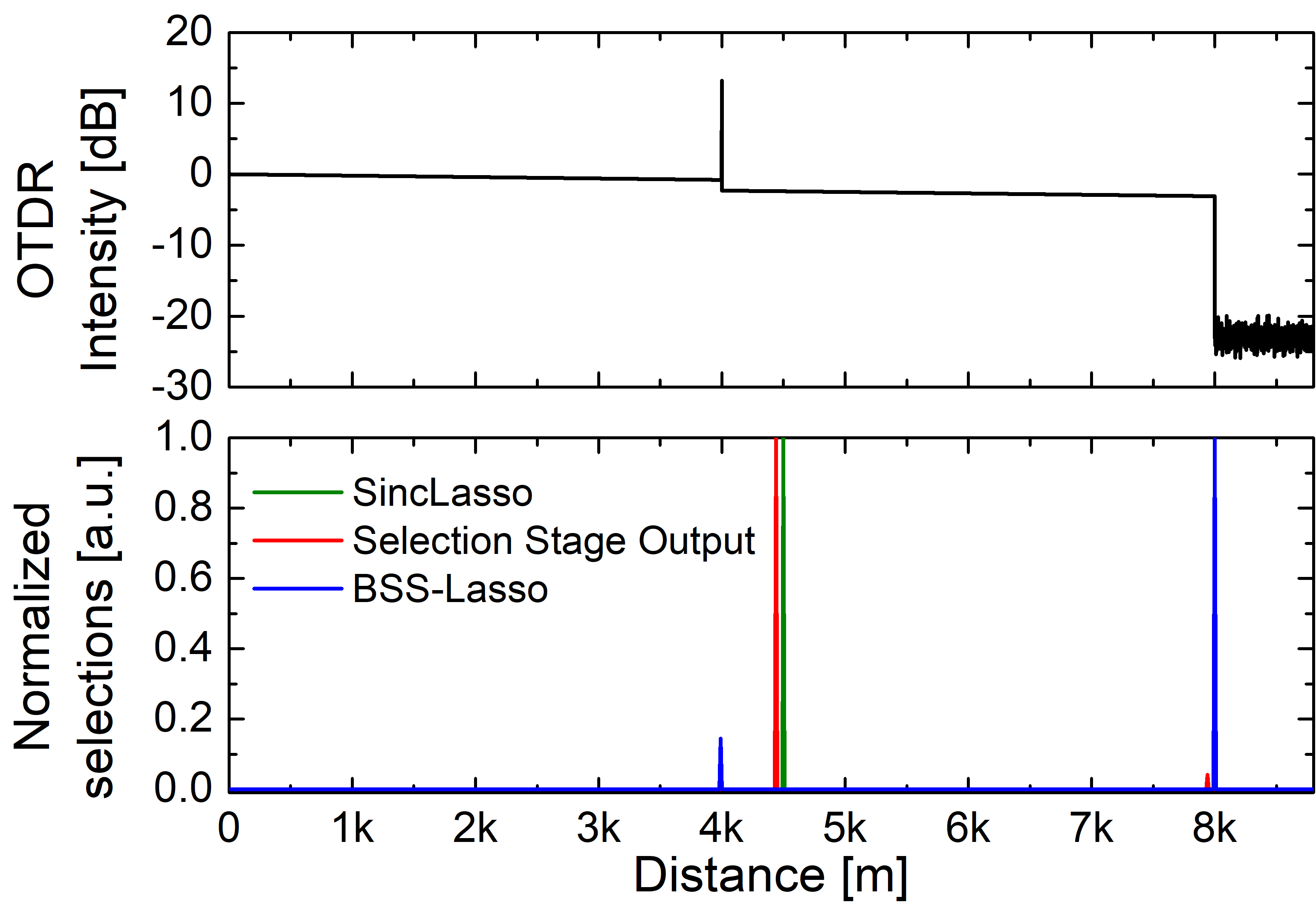}
		\caption{Comparison of the normalized selections resulting from SincLasso, BSS-Lasso up to the selection stage and complete BSS-Lasso for an 8 km fiber link with a reflective fault at 4 km and a non-reflective fault at 8 km.}
		\label{fig:sinc_3p_comparison}
	\end{figure}

    The rationale behind the correction stage is that, if a reflection was found at a given position, it generally indicates the presence of a fault at that location. Additionally, if Algorithm \ref{alg:correction_stage}, due to the reduced penalty, selects at the correct location a fault that was previously shifted, this often causes the algorithm to reject its previous incorrect position because of the $\ell_1$-norm penalty, thus correcting the shifting phenomenon. For illustrative purposes, Fig. \ref{fig:sinc_3p_comparison} depicts the monitoring accuracy of the SincLasso, BSS-Lasso up to the selection stage and complete BSS-Lasso for an 8 km fiber link. Note that, for the reflective fault at 4 km, both the SincLasso and the BSS-Lasso up to the selection stage incurred in an error of over 400 m, while the complete BSS-Lasso presented a 10 m error due to the correction stage.
    
	\begin{remark}
    	In this work, the proposed algorithm has up to three overall Lasso iterations. Although additional iterations might be justifiable to further improve the quality of the method, there is a lack of a consistent stopping criterion in order to avoid cycling. In fact, two overall iterations are sufficient for the existence of a correction stage. However, it was empirically observed that a third one, which acts as a refinement step, significantly improves the precision of the method. Furthermore, the benefits of additional iterations (more than three) were observed to be marginal, non-existent or even negative. For these reasons, the number of iterations was fixed at three in Algorithm \ref{alg:correction_stage}, which results in a computationally effective procedure.
	\end{remark}
    
	\subsection{Treatment stage of the BSS-Lasso}
	
\begin{figure}[h]
		\centering
		\includegraphics[width=0.7\columnwidth]{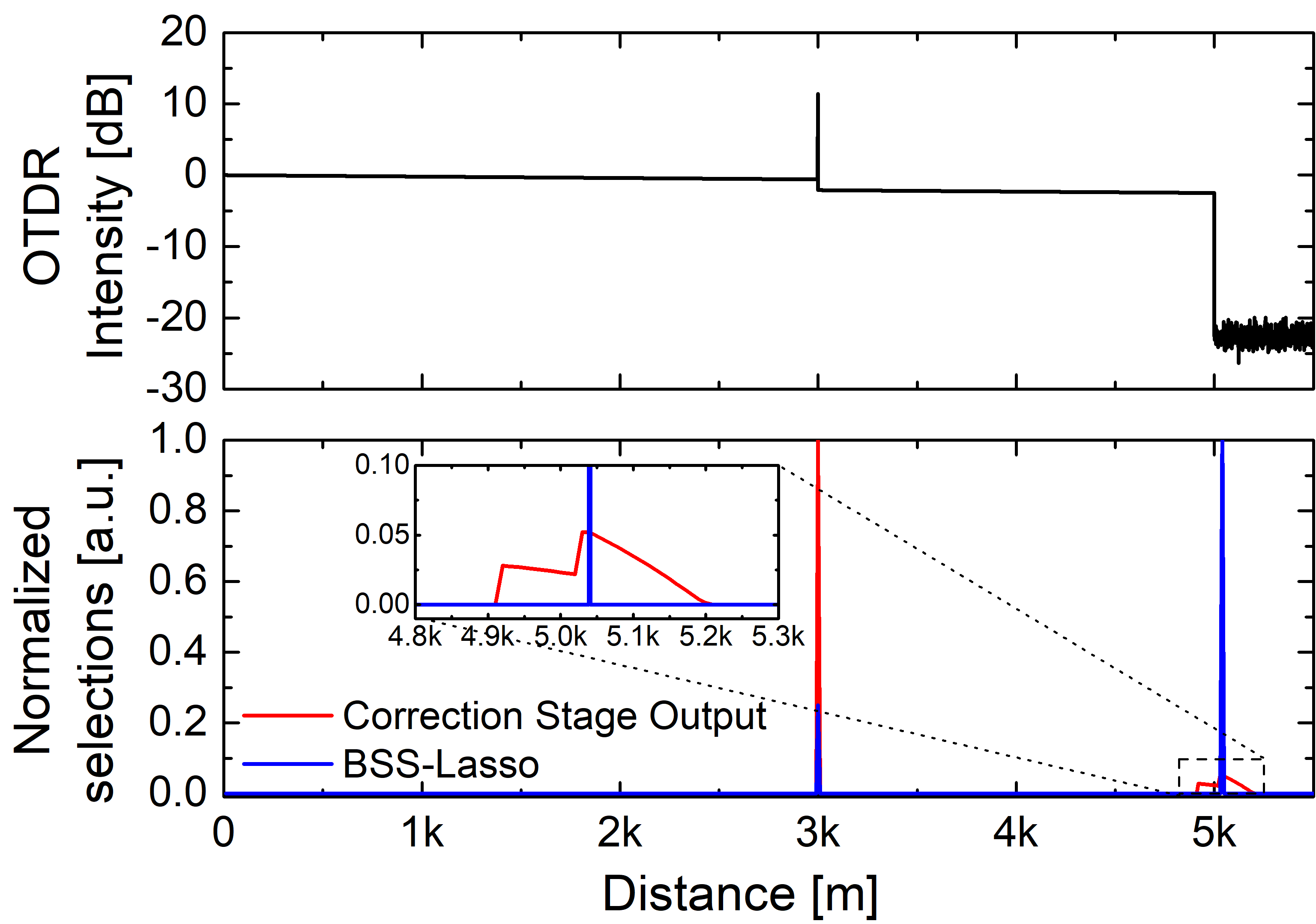}
		\caption{Time-domain profile in the top panel and BSS-Lasso selections in the bottom panel for a 5 km fiber link with a reflective event at 3 km and a non-reflective event at 5 km. The reflective event results in a single selection, while the non-reflective one induces a cluster of selections.}
		\label{fig:clusters}
	\end{figure}

    Ultimately, Algorithm \ref{alg:correction_stage} outputs a $2q$-dimensional vector indicating the fault locations. However, in many cases, the result is not presented as a straightforward handful of singular selections, but rather as a set, or cluster, of event positions. Although the appearance of such clusters may have its roots in the limited spatial resolution of the monitoring technique, the Lasso is known to produce clusters whenever the over-complete dictionary variables present high levels of correlation \cite{Hastie2009_ElemStatLearn}. In fact, depending on the fiber discretization, adjacent columns of the explanatory matrix $\mathbf{M}$ represent close enough positions such that their induced signals are almost indistinguishable, thus strongly correlated. As a consequence, the selection and correction stages oftentimes output multiple selections around the real fault position as a cluster (see Fig. \ref{fig:clusters} for an illustrative example containing an output from the correction stage and the final BSS-Lasso result) and it might be useful to refine the result in order to obtain more accurate selections to better assist the network operator.
	
    A handful of techniques appears in technical literature to handle correlated explanatory variables within the Lasso framework. The commonly used approach performs an \textit{a priori} clustering of the correlated explanatory variables before solving the Lasso problem \cite{Buhlmann2013_CorrelatedLASSO}. However, such \textit{a priori} clustering is not suited for the particular application of this work, since all explanatory variables are sequentially correlated to its adjacent neighbors with respect to the employed distance grid in matrix $\mathbf{M}$. To overcome this issue, an alternative is to narrow down the clusters of selections into single selections after the Lasso processing. From a technical point-of-view, a cluster of selections represent a degenerate projection of $\mathbf{y}$ onto the column space of $\mathbf{M}$, i.e, one possible way to express this projection in terms of the column vectors of $\mathbf{M}$. As a consequence, it is likely that the real fault is among the cluster positions and closer to the selections with higher magnitude. Therefore, one can interpret each cluster of selections as a probability density for the respective true fault position. In this context, several approaches can be applied to directly narrow down the clusters of selections.
    
    Firstly, since the clusters can be interpreted as probability densities, a computationally efficient approach is to estimate the true fault position by applying a weighted average of the positions within the clusters, where the weights are given by the selection magnitudes. Another possibility is to consider all possible combinations of positions with a single selection per cluster and choose the one which outputs the least square model fitting error (minimal $\ell_2$ norm). This approach can be implemented by either a combinatorial set of ordinary least squares computations or via Mixed Integer Quadratic Programming (MIQP) \cite{Bertsimas2016_BestSubSelection, Bertsimas2016_AlgoApprochForLinear}. Despite showing the best empirical results, in theory, it is affected by the curse of dimensionality and can be computationally burdening, i.e., instances with wide and/or several clusters (e.g., a fiber with numerous faults) might be intractable in reasonable computational time. Nevertheless, it should be emphasized that, for most practical cases, this approach can be computed in the order of seconds. Finally, a viable alternative to the combinatorial least squares for fibers with multiple events is the least absolute error (minimal $\ell_1$ norm), which can be solved using Mixed Integer Linear Programming (MILP) algorithms. Although still a combinatorial procedure due to their integer nature, MILP problems are widely recognized to be more computationally efficient to be solved than MIQP.
    
    Due to its better empirical performance, in this work, the treatment of the clusters of selections is performed using least squares. It should be highlighted that this choice is more consistent with the proposed BSS-Lasso methodology, since the Lasso already uses the $\ell_2$ norm for the model fitting (see the first term of the objective function in \eqref{eq:lasso}). In fact, this choice of \textit{ex-post} cluster treatment can be seen as an instance of the Lasso's original design \cite{tibshirani1996regression}, in which the $\ell_1$ regularization term (second term of the objective function in \eqref{eq:lasso}) is replaced by the semi-norm $\ell_0$ and written as a constraint bounded by 1 for each cluster. More precisely, let $\{\mathcal{C}_{i}\}_{i = 1}^{C}$ be the family of $C$ clusters resultant from the BSS-Lasso and $\overline{\mathbf{M}} = \big[\big\{\mathbf{M}_{j}\}_{j \in \mathcal{C}_{i}}\big\} \big]_{i = 1}^{C}$, the explanatory matrix restricted to the position indexes within each cluster. Then, the cluster treatment procedure can be formulated as the following mathematical programming problem (\cite{Bertsimas2016_BestSubSelection, Bertsimas2016_AlgoApprochForLinear} are referred to for a wider discussion and efficient formulations for this problem):
    \begin{align}
		\min_{\boldsymbol{\beta}} \left\{ \big\|\mathbf{y} - \overline{\mathbf{M}} \boldsymbol{\beta} \big\|^{2}_{2} ~ \middle| ~ \begin{array}{l r}
        	\big\|\overline{\boldsymbol{\beta}}_{i}\big\|_{0} \leq 1, & \forall ~ i \in \mathcal{C}; \\[0.50em]
            \boldsymbol{\beta} \geq \boldsymbol{0}; &
         \end{array} \right\}, \label{eq:Reducedlasso}
	\end{align}
    where $\overline{\boldsymbol{\beta}}_{i} = \{\beta_{j}\}_{j \in \mathcal{C}_{i}}$. Therefore, in this context, the original $\ell_0$-Lasso is solved, but avoiding its high combinatorial nature since the $\ell_0$ search is performed in a drastically reduced space when compared to the original problem. The treatment stage of the BSS-Lasso is thus presented in Algorithm \ref{alg:treatment_stage}.

    \begin{algorithm}[H]
		\caption{BSS-Lasso -- Treatment stage}
		\begin{algorithmic}
        \label{alg:treatment_stage}
        	\STATE Let $\hat{\boldsymbol{\beta}}^{(3)}$ be the correction stage output.
            \STATE $i \gets 1$, $\mathcal{C}_{i} \gets \{\}$, $C \gets i$.
            \FOR{$j = 1$ \TO $q$}
            	\STATE If $\hat{\beta}^{(3)}_{j} > 0 \implies \mathcal{C}_{i} \gets \mathcal{C}_{i} \cup \{j\}$
                \STATE Else $i \gets i + 1; \hspace{0.20cm} \mathcal{C}_{i} \gets \{\}$; $C \gets i$
            \ENDFOR
            \STATE Construct $\overline{\mathbf{M}} = \big[\big\{\mathbf{M}_{j}\}_{j \in \mathcal{C}_{i}}\big\} \big]_{i = 1}^{C}$
            \STATE Solve $\ell_0$-Lasso \eqref{eq:Reducedlasso} and \textbf{return} its optimal decision vector
		\end{algorithmic}
	\end{algorithm}
    
    For future reference, BSS-Lasso will be identified as the sequential computing of Algorithms \ref{alg:selection_stage}--\ref{alg:treatment_stage}. % LASSO, High-dimension problems, and the BSS-Lasso
    
    %------------------------------------------------
	
    \section{Experimental Validation} \label{sec:exp_results}
\begin{figure}[h]
		\centering
		\includegraphics[width=0.7\columnwidth]{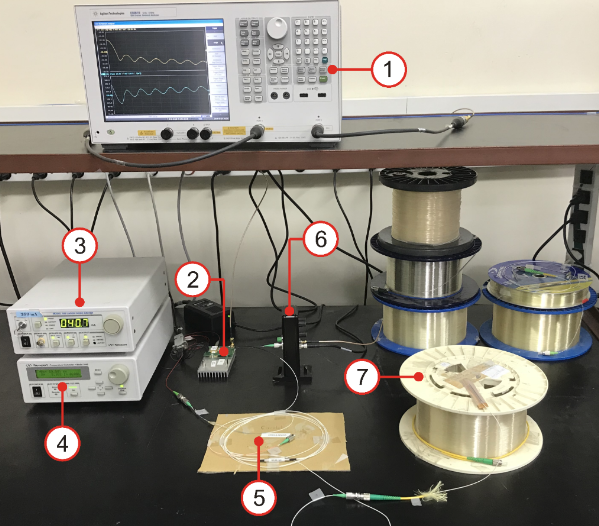}
		\caption{Photo of the experimental setup. 1) Network Analyzer; 2) Distributed Feedback-Laser Diode; 3) Laser Bias Source; 4) Temperature Controller Driver; 5) Optical Circulator; 6) Photodetector; 7) Fiber under test.}
		\label{fig:exp_photo}
	\end{figure}

    Validation of the BSS-Lasso in a real environment consists in comparing its estimated events positions with the reference events positions determined using a standard OTDR device for different fibers. Practically, the limitation on the number of experimentally tested links is determined by the availability of fibers with different lengths in the laboratory and the possibility of connecting these fibers to form links with different number of events and different lengths. Six fiber links were available, for which the profiles have been individually measured; combinations of these six fibers two-by-two, to compose two-event links, produced 30 more examples. Prior to measuring the fibers, however, a few experimental parameters concerning the data acquisition and the minimum monitoring signal modulation power must be defined. A picture of the experimental setup as assembled in the laboratory for all these tests is presented in Fig. \ref{fig:exp_photo}.

\subsection{Experimental parameters characterization}

As depicted in the experimental setup of Fig. \ref{fig:exp_setup} and shown in Fig. \ref{fig:exp_photo}, the optoelectric conversion and signal amplification is performed by a photodetector, which is composed mainly by a photodiode and two amplifiers. From an optical perspective, as long as the backscattered optical signal reaching the photodiode carries power greater than the NEP, the output electrical signal will carry information about the fiber. In the same way, but from an electrical point of view, the signal reaching the NA must be greater than its NEP ($N_{na}$), so that the measured frequency profile translates the fiber's characteristics. Since an important characteristic of the Baseband Subcarrier Sweep monitoring technique is its coexistence with data transmission in a transmitter-embedded configuration, as discussed in \cite{amaral2017low}, it is paramount that a balance between monitoring signal power and the capacity of meeting the previously mentioned power requirements, in both the photodetector and the NA, is found. In this subsection, this balance will be analyzed and the experimental parameters used throughout the BSS-Lasso validation will be presented.

Analysis of the optical signal-to-noise ratio (SNR) requires knowledge of a few parameters, all of which are summarized in the upper part of Table \ref{tab:expParam}. The laser diode's bias current is set at 70 mA, yielding an output optical power of 4 dBm; this allows for a 52 mA excursion inside its linear region. An insertion loss of approximately 1 dB in the optical circulator (port 1 $\rightarrow$  port 2) sets the input optical power to the fiber under test (FUT) to $\sim\!3$ dBm. Thus, considering a Rayleigh backscattered coefficient $C$ of -72 dB/m \cite{DericksonBOOK1998}, an attenuation coefficient $\alpha$ of 0.2 dB/km, a 6372 m optical fiber, and another 1 dB of insertion loss in the optical circulator (port 2 $\rightarrow$  port 3), the optical power arriving at the photodiode in the steady-state regime would be $-33.20$ dBm (or 477 nW). This value has been experimentally verified to be $-34.02$ dBm. The employed photodetector exhibits a NEP of 200 pW in a full bandwidth condition \cite{detDatasheet}, so a comfortable 33 dB SNR at the photodiode is ensured. Note, however, that this SNR represents the total optical power arriving at the photodiode, including the DC component. As will be discussed presently, a 10\% portion of the laser's full modulation depth is occupied by the monitoring signal, so the SNR for the signal of interest is approximately 23 dB.

Electrical signal analysis begins right after the opto-electrical conversion of the photodiode: although the two amplification stages of the photodetector exhibit noise figures of their own and also amplify noise coming from the photodiode, the optical SNR is sufficiently high so that one can consider the output signal of the photodetector to maintain the same SNR. Thus, taking into account the received optical power of 477 nW and the photodetector parameters (responsivity, transimpedance gain, and the 2$^\textrm{nd}$ stage voltage gain) and also the portion of the full modulation depth occupied by the monitoring signal, all shown in Table \ref{tab:expParam}, it is possible to calculate an amplitude of 298.37 mV for the input signal entering port 2 of the NA at low frequencies. According to the mathematical model and the experimental results, the amplitude of the electrical signal has an inverse dependence with the frequency and, as the latter increases, the former is expected to dramatically decrease. The value for the low-frequency measurement, however, has been experimentally measured to be 303.57 mV, which is far above $N_{na}$ ($-$95 dBm \cite{NADatasheet}), setting the electrical SNR at 47.32 dB and validating both the optical and electrical signal analysis.

To address the SNR reduction as the frequency increases, which can translate into up to 30 dB reduction at the maximum swept frequency, an averaging process is employed. This consists of acquiring several traces inside the same interval and taking the arithmetic average, a resource already available in the employed NA. At the same time, since the frequency-domain profile is determined in a steady-state regime, the intermediate frequency bandwidth of the NA frequency beat detector is set to a low value (150 Hz) translating into a considerable long time-constant of 6.64 s. This long time constant translates into an intrinsic averaging of the measured amplitude and phase of the input signal, which, by itself, diminishes noise contributions to the measurement, especially at higher frequencies, and alleviates the number of samples for the average procedure. Therefore, in order to reach a compromise between faster data acquisition and negligible noise contribution, a total of 10 samples has been set as default to the averaging procedure; the total data acquisition process amounts to less than two minutes.

\begin{table}[h]
\centering
\caption{Experimental Parameters for Data Acquisition}
\label{tab:expParam}
\begin{tabular}{|c|l|l|}
\hline
\multicolumn{3}{|c|}{\textbf{Experimental Parameters}}                            \\ \hline
\multirow{4}{*}{Optical}    & Fiber input power ($P_{0}$) & 3 dBm \Tstrut           \\ \cline{2-3} 
                            & Fiber length              & 6372 m \Tstrut          \\ \cline{2-3} 
                            & Circulator loss           & 1 dB  \Tstrut           \\ \cline{2-3} 
                            & Fiber attenuation  ($\alpha$)            & 0.2 dB/km \Tstrut       \\ \hline
Opto/Electric               & PD responsivity @ 1550 nm ($R$) & 1 A/W     \Tstrut       \\ \hline
\multirow{5}{*}{Electrical} & PD transimpedance gain    & 626 V/A   \Tstrut       \\ \cline{2-3} 
                            & PD 2$^{\textrm{nd}}$ stage voltage gain         & 1x10$^4$ V/V  \Tstrut \\ \cline{2-3} 
                            & Laser Bias Source (LBS)	& 70 mA    \Tstrut        \\ \cline{2-3} 
                            & NA modulation power     & 0 dBm    \Tstrut        \\ \cline{2-3} 
                            & Modulation depth ($m$)        & 10\%     \Tstrut        \\ \hline
\end{tabular}
\end{table}
    
\begin{figure}[h]
        \centering
        \includegraphics[width=0.7\columnwidth]{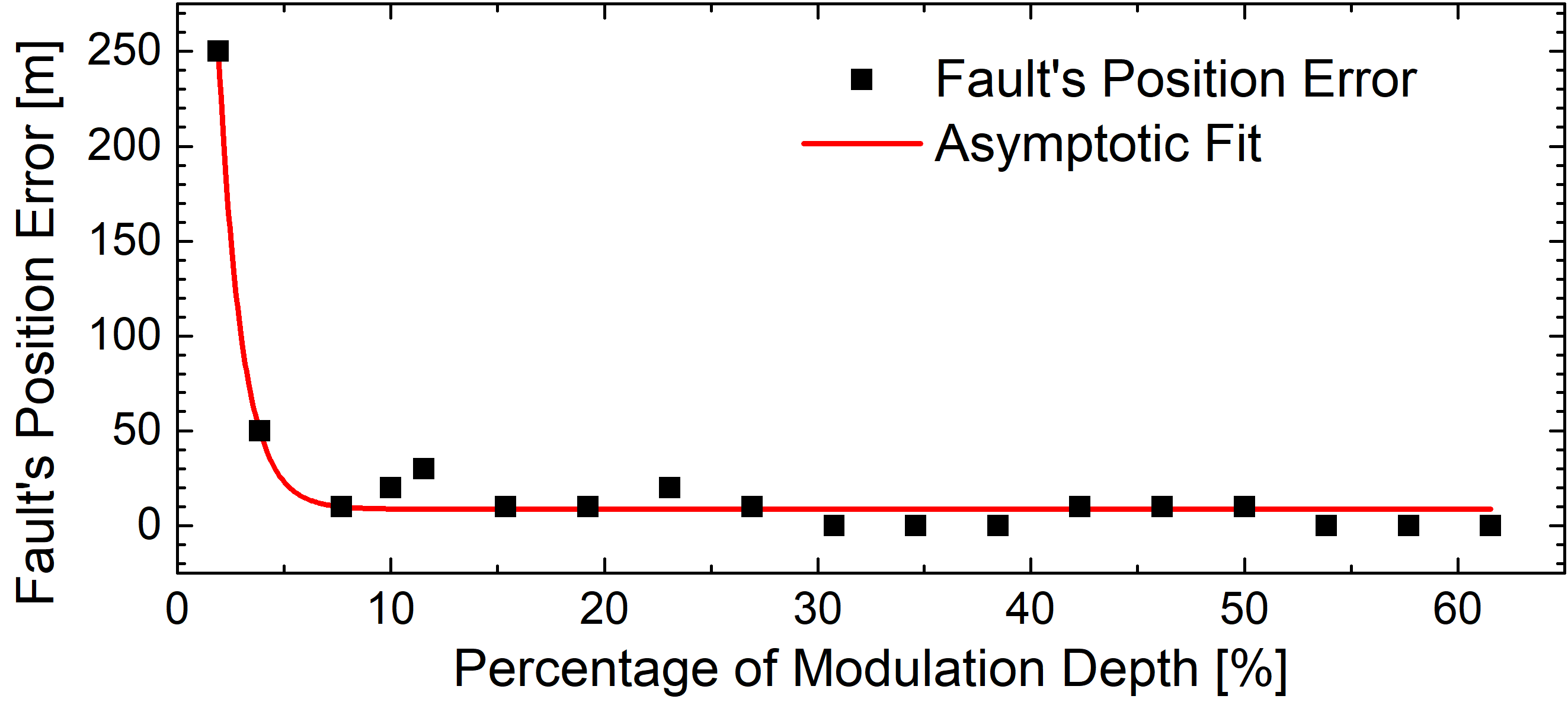}
        \caption{Impact of the low-frequency monitoring signal's amplitude -- written in terms of percents of the laser's full modulation depth -- on the BSS-Lasso event position estimation. The 5\% mark represents a clear lower bound on the amplitude of the monitoring signal so that reliable results are obtained.}
        \label{fig:elec_pow_analysis}
    \end{figure}

    As mentioned in the previous analysis, the portion of the laser's full modulation depth occupied by the monitoring signal was set to 10\% as the balance between minimal monitoring power that allows for accurate fiber characterization. This value immediately impacts on the capacity of concurrent data transmission in a direct modulation scheme and experimentally characterizes the BSS-Lasso technique in the context of monitoring and data coexistence. To validate this parameter, the signal's amplitude entering the modulation port of the laser was varied from 50 mVpp to 1.6Vpp, which corresponds, approximately, to 2\% and 62\% of the laser's full modulation depth, respectively. The BSS-Lasso was tested for each signal amplitude in the same fiber testbed (with parameters described in Table \ref{tab:expParam}), with the frequency-domain fiber profile being determined with the BSS setup, and errors between the real positions of events and the ones obtained from the technique being evaluated based on the output of the BSS-Lasso. The experimental result is presented in Fig. \ref{fig:elec_pow_analysis}.
    
    The results validate the experimental parameters presented in Table \ref{tab:expParam} and show that, for monitoring signals with amplitude greater than 10\% of the modulation depth, position estimates are within a 30 m distance error from the actual positions. When the monitoring signal's amplitude falls below 5\% of the modulation depth, on the other hand, the distance error rapidly increases. This indicates that the BSS-Lasso can be implemented to continually monitor a fiber link while occupying only a tenth of the laser's full modulation depth, thus making available a considerable portion of the laser's electro-optical transfer function for the purpose of data transmission. Furthermore, we note that the precision plateau attained at 10\% of the full modulation depth is associated to a limitation of the spatial resolution of the technique, which is related to the maximum monitoring frequency. All the experimental results presented in this paper have been acquired using the parameters described in Table \ref{tab:expParam}.

\subsection{Experimental results}

\begin{figure}[h]
		\centering
		\includegraphics[width=0.6\columnwidth]{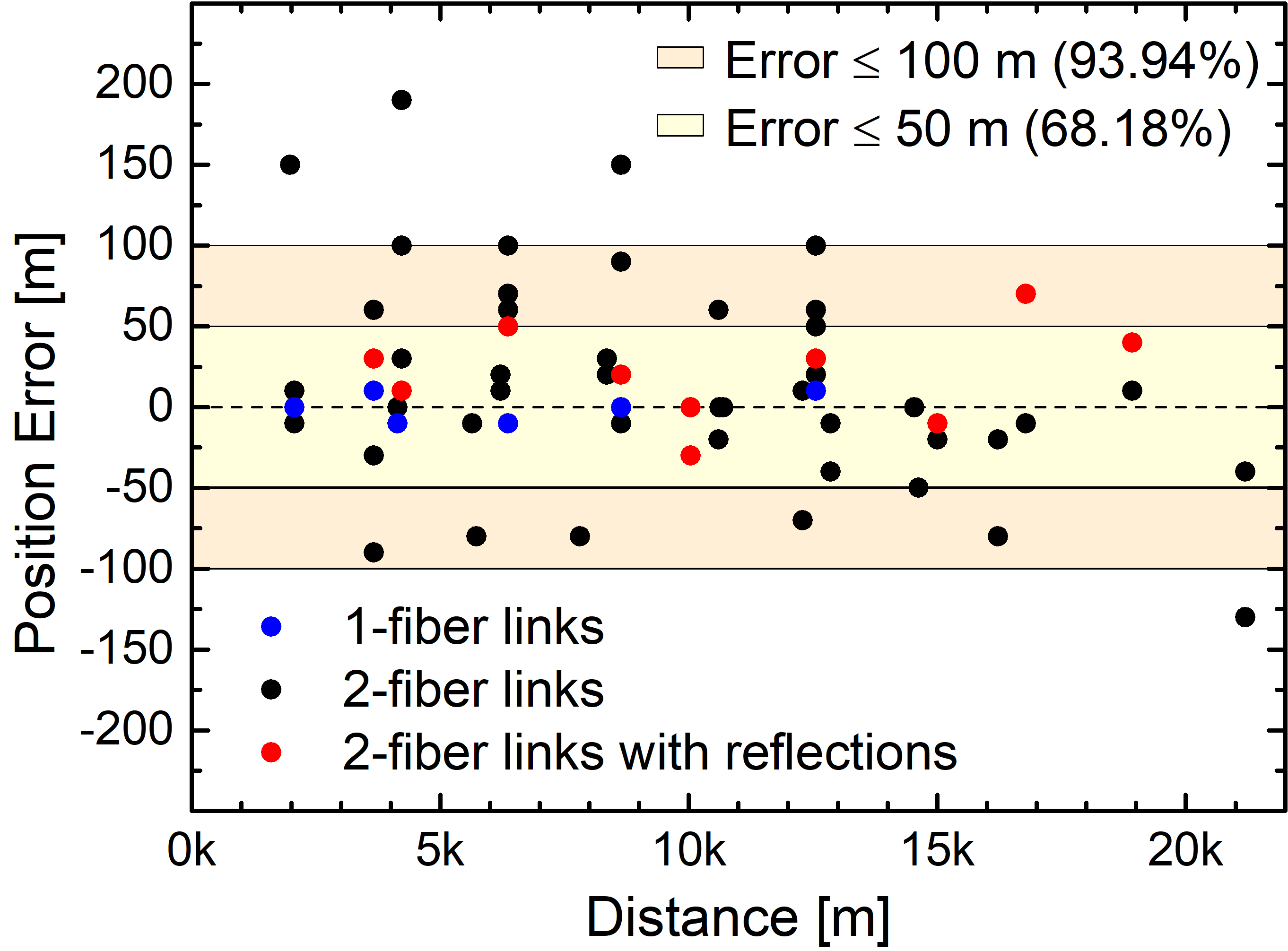}
		\caption{Distribution of position estimation errors for all the experimentally measured fiber profiles. Apart from a few outliers, the positions are within a 100 meters error interval from the actual real fault positions.}
		\label{fig:exp_result_scatter}
	\end{figure}

In Fig. \ref{fig:exp_result_scatter}, the event position estimation error of the BSS-Lasso is compared with the reference results of a standard OTDR device and presented in the form of a scatter plot. Two error thresholds are defined in the scatter; the first one delimits a low-error interval, where the results are deemed to be very close to the reference, in which almost 70\% of the results have fallen; the second delimits results that are fairly close to the reference position, and holds over 93\% of the total results. A few outliers can be observed with errors above 100 m. The error distribution, mainly concentrated around the zero error point and within the 0-50 m error interval attests the prowess of the BSS-Lasso in characterizing the assembled experimental fiber links.

    Interesting observations can be made from the results of Fig. \ref{fig:exp_result_scatter}. First of all, it is clear that the estimation error tends to be closer to zero for single-fiber links. As will also be discussed in Section \ref{sec:sim_results}, the estimation is hindered by the amount of events present in the fiber link. Secondly, it is noticeable that the presence of reflective events induces a better estimation of the event's positions, as all but one of them fell within a 50 m error even with a higher number of events (two non-reflective events and one reflective event). This characteristic, which will also be studied in Section \ref{sec:sim_results}, indicates that the modeling of reflective events paired with the correction stage increases the robustness of the estimation.
    
   Up to this point, accurate determination of the faults positions has been demonstrated using the BSS-Lasso. However, determination of the fault magnitudes is of utmost importance for full characterization of the optical fiber link, and can be performed following Eq. \eqref{eq:magCoeff}. Nevertheless, even though the coefficients $\phi_b$, determined as a product of the BSS-Lasso, allow for accurate estimation of the BSS frequency-domain profile, they include the bias naturally induced by the Lasso \cite{tibshirani1996regression}. Moreover, the relationship between the $\phi_b$ and the actual fault magnitudes $\xi_b$ (Eq. \eqref{eq:magCoeff}) is extremely non-linear, so eventual deviations in the former induce severely imprecise results for the latter.
   
   To ensure correct estimation of the fault magnitudes, a \textit{reconstruction} procedure of the time-domain profile of the fiber, that does not depend on the amplitudes determined by the BSS-Lasso, is proposed. It is as follows: (i) utilizing the event position estimates, fiber link profiles are created in the form of Eq. \eqref{eq:PzHeaviside} for which the respective non-reflective magnitudes $\xi_b$ are sorted within a predetermined interval of \text{[0, 5]} dB in a sequence of two steps, a coarser 0.5 dB, and then a finer 0.1 dB; (ii) the reflective magnitudes $\theta_r$ are sorted within a predetermined interval of \text{[0, 20]} dB in steps of 2 dB; (iii) from the created $P\left(z\right)$, frequency-domain profiles $S\left(f\right)$ are calculated using Eq. \eqref{eq:Sf_IntForm} and then compared to the estimated frequency-domain profile given by the BSS-Lasso; (iv) the created $S\left(f\right)$ are compared with the BSS-Lasso estimated profile using the $\ell_2$ error norm; and (v) the best reconstructed fiber profile according to this metric is defined as the reconstructed fiber profile. Presented in Table \ref{tab:magCompare}, for four experimental fiber links, are: the real fault magnitudes, calculated using the OTDR profile measured by a standard OTDR device; the magnitudes calculated using Eq. \eqref{eq:magCoeff} and the BSS-Lasso coefficients; and the magnitudes determined through the reconstruction procedure.
   
\begin{table}[h]
\centering
\caption{Fault Magnitude Comparison}
\label{tab:magCompare}
\begin{tabular}{|c|c|c|}
\hline
Real & Calculated & Reconstructed\\
Magnitudes [dB] & Magnitudes [dB] & Magnitudes [dB]\\
\hline
1.9, 23.0 & 2.77, $\infty$ & 1.6, 22.0 \Tstrut\\
2.0, 20.7 & 1.79, $\infty$ & 2.0, 22.0 \Tstrut\\
2.9, 17.5 & 3.53, $\infty$ & 3.0, 22.0 \Tstrut\\
0.9, 22.2 & 1.92, $\infty$ & 1.0, 22.0 \Tstrut\\
\hline
\end{tabular}
\end{table}
   
\begin{figure}[h]
		\centering
		\includegraphics[width=0.7\columnwidth]{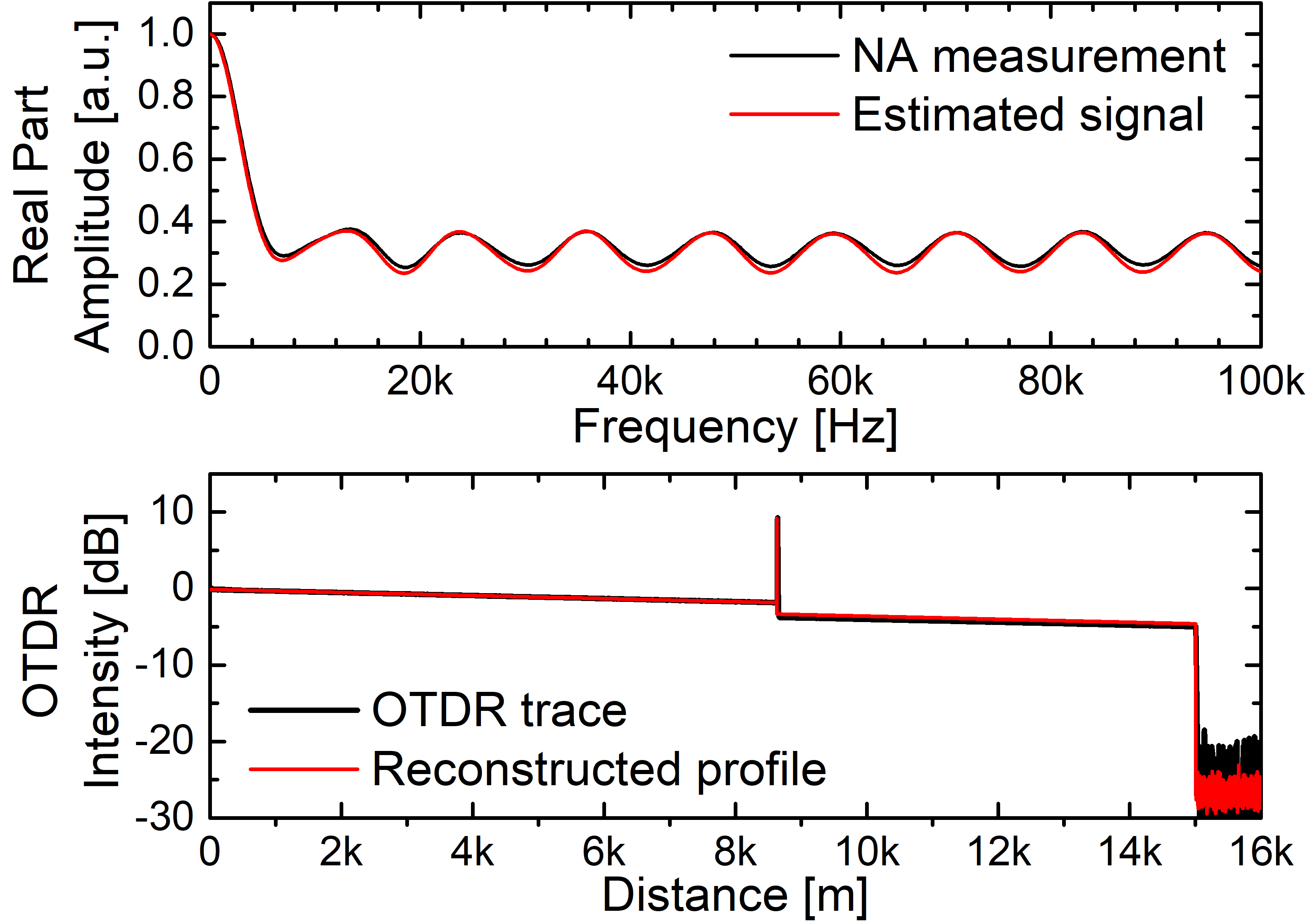}
		\caption{Time-domain profile reconstruction based on the results of the BSS-Lasso. In the upper panel, the estimated frequency-domain profile of the BSS-Lasso is plotted against the $S\left(f\right)$ that corresponded to the minimum $\ell_2$ error norm during the reconstruction step. In the lower panel, the original OTDR profile acquired with a standard OTDR device is depicted along with the reconstructed profile.}
		\label{fig:reconstruction}
	\end{figure}

   In other words, the reconstruction procedure is equivalent to the second analysis path presented in Section \ref{Sec:Eq_TimeFreq}, but using artificially created $P\left(z\right)$ based on the results of the BSS-Lasso. The result, as can be perceived from Table \ref{tab:magCompare}, is a much more accurate estimate of the fault magnitudes. Finally, an example of the reconstruction procedure for the first fiber link of Table \ref{tab:magCompare}, i.e., the artificially generated $P\left(z\right)$ that translates into $S\left(f\right)$ that best approximates the estimated frequency-domain signal given by the BSS-Lasso, is depicted in Fig. \ref{fig:reconstruction}.
    
     % Experimental Results
    
    %------------------------------------------------
    
	\section{Simulation Results} \label{sec:sim_results}
    \newcolumntype{L}[1]{>{\raggedright\arraybackslash}p{#1}}
\newcolumntype{C}[1]{>{\centering\arraybackslash}p{#1}}
\newcolumntype{R}[1]{>{\raggedleft\arraybackslash}p{#1}}
\begin{table*}[h!]
	\renewcommand{\arraystretch}{1.20}
	\centering
  \footnotesize
  \caption{Selection errors for each set of simulated links}
  \label{tab:errors}
  \begin{tabular}{C{1.20cm} | C{1.30cm} C{1.30cm} C{1.30cm} | C{1.30cm} C{1.30cm} C{1.30cm} | C{1.30cm} C{1.30cm} C{1.30cm}|}
    \cline{2-10}
     & \multicolumn{3}{c|}{1 fault} & \multicolumn{3}{c|}{2 faults} & \multicolumn{3}{c|}{3 faults} \\ \hline
    \multicolumn{1}{|c|}{Error [m]} & SincLasso & BSS-1 & BSS-Lasso & SincLasso & BSS-1 & BSS-Lasso & SincLasso & BSS-1 & BSS-Lasso \\ \hline
    \multicolumn{1}{|c|}{{[}0, 50{]}} & 52.20\% & 52.20\% & 89.80\% & 49.60\% & 49.45\% & 81.50\% & 50.70\% & 50.77\% & 77.50\% \\
    \multicolumn{1}{|c|}{(50, 100{]}} & 1.60\% & 1.60\% & 1.60\% & 1.20\% & 1.45\% & 2.95\% & 2.60\% & 2.87\% & 4.03\% \\
    \multicolumn{1}{|c|}{(100, 200{]}} & 11.80\% & 11.90\% & 7.20\% & 4.40\% & 5.85\% & 5.65\% & 3.33\% & 4.93\% & 4.30\% \\
    \multicolumn{1}{|c|}{(200, $\infty$)} & 34.40\% & 34.30\% & 1.40\% & 44.80\% & 43.25\% & 9.90\% & 43.37\% & 41.43\% & 14.17\% \\ \hline
  \end{tabular}
\end{table*}

\begin{table*}[h!]
	\renewcommand{\arraystretch}{1.20}
    \centering
    \footnotesize
    \caption{Contingency table ($\pm$ 50 m)}
    \label{tab:contingency}
    \begin{tabular}{c|c|c|c||c|c|c|}
        \cline{2-7}
        & \multicolumn{3}{c||}{Fault Present} & \multicolumn{3}{c|}{Fault Absent} \\ \cline{2-7}
        & \multicolumn{1}{c|}{SincLasso} & \multicolumn{1}{c|}{BSS-1} & \multicolumn{1}{c||}{BSS-Lasso} & \multicolumn{1}{c|}{SincLasso} & \multicolumn{1}{c|}{BSS-1} & \multicolumn{1}{c|}{BSS-Lasso} \\ \hline
        \multicolumn{1}{|c|}{\multirow{2}{*}{Fault Found}} & True Positives & True Positives & True Positives & False Positives & False Positives & False Positives \\
        \multicolumn{1}{|c|}{} & 3035 & 3034 & 4853 & 2808 & 3251 & 1379 \\ \cline{1-7}
        \multicolumn{1}{|c|}{\multirow{2}{*}{Fault Neglected}} & False Negatives & False Negatives & False Negatives & True Negatives & True Negatives & True Negatives \\
        \multicolumn{1}{|c|}{} & 2945 & 2966 & 1147 & 2959557 & 2958128 & 2957685 \\ \cline{1-7}
        \multicolumn{1}{|c|}{\multirow{2}{*}{Measures}} & Sensitivity & Sensitivity & Sensitivity & Specificity & Specificity & Specificity \\
        \multicolumn{1}{|c|}{} & 50.58\% & 50.57\% & 80.88\% & 99.91\% & 99.89\% & 99.95\% \\ \hline
    \end{tabular}
\end{table*}

In order to illustrate the robustness of the proposed monitoring methodology, in this section, extensive computational tests are conducted on the BSS-Lasso. To do so, a large-scale test bench was created: a total of three sets of 1000 fiber links each containing, respectively, one, two, and three faults were randomly generated based on the following steps:
    \begin{enumerate}
    	\item Fiber lengths are sampled evenly within $L \in [2, 15]$ km.
        \item Given the fiber length sampled in Step 1, the fault locations are also sampled evenly within $[2, L]$, with a  mandatory fault at the end of the fiber.
        \item In order to sample reflective events, a 50\% chance of having an associated reflection is attributed to each fault sampled in Step 2.
        \item The magnitudes of the events are then randomly chosen, with faults ranging evenly between $[1, 5]$ dB and reflections up to 20 dB.
        \item The time-domain profile $P\pars{z}$ of the fiber link is constructed using Eq. \eqref{eq:PzHeaviside} and the parameters sampled in Steps 1--4.
        \item The methodology presented in Section \ref{sec:dataAcq_Model} is utilized to obtain the frequency-domain profile $S\pars{f}$ from $P\pars{z}$, using Eq. \eqref{eq:Sf_IntForm} with a set of frequencies ranging within [100, 100000] Hz discretized in 100 Hz.
    \end{enumerate}
    For reproducibility purposes, the complete test bench of fiber links used in this section is available in \cite{TestBenchIEEE}. All tests were conducted in Julia language with an Intel Core i7-490K CPU at 4.00 GHz and 32 GB of RAM memory. The simulation parameters used throughout the testbench were: 10 m length discretization; reduced penalty factor $\gamma = 0.5$; and sensitivity threshold $\epsilon = 0.05$. For the sake of comparison, the performance of the BSS-based monitoring technology presented in \cite{amaral2017multiple}, known as \textit{SincLasso}, was also evaluated for the same test bench. Furthermore, in order to quantify the contribution of the correction stage described in Algorithm \ref{alg:correction_stage}, the results of the BSS-Lasso up to the selection stage (named \textit{BSS-1} for presentation purposes) are also studied.
    
    The test results are presented in two separate tables, for more rigorous analysis. In Table \ref{tab:errors}, the errors are stratified among four distance intervals and the respective percentage of position estimates within each interval is evaluated for the three techniques. Table \ref{tab:contingency} presents the so-called contingency table. The idea is to perform a binary classification of \textit{fault/no fault} events based on the output of the techniques. More precisely, in the contingency table, True Positives are accounted by estimations within the low-error interval, i.e., when the selection error is 50 m or lower. Similarly, False Positives are estimations outside the low-error interval. It should be highlighted that Table \ref{tab:contingency} was constructed by combining the results of the three sets of fiber links, totaling 6000 events.
    
    The first conclusion based on the analysis of both tables is the clear dominance of the BSS-Lasso over the other two techniques; the BSS-Lasso achieves over 80\% of fault estimates within the low-error interval while the other techniques barely reach 50\%. A second observation is that the number of events clearly impacts the accuracy of the methodology, as it diminishes from 89.80\% for the set of fiber links with a single event to 77.50\% for the set with three events. The increase in estimation difficulty as the number of events grows larger is expected, as the method needs to distinguish more components that are compounded in the signal. It is also striking that the selection stage of the BSS-Lasso has performance extremely similar to the SincLasso, indicating that the inclusion of the reflections in the dictionary of the Lasso is not sufficient to accurately detect reflective events. Furthermore, the near 50\% accuracy limit of the SincLasso shares a strong correlation with the definition of Step 3 in the test bench creation protocol, since the probability of a reflective event was set to 50\%. This result translates the lack of precision induced by a reflection and indicates that the BSS-Lasso not only deals with such reflective events but uses their presence to optimize the estimation through the correction stage.
    
	The BSS-Lasso also excels when analyzing the contingency table, with a number of False Negatives and False Positives less than half of those of the other techniques. Even though a finer analysis of the specificity\footnote{$\text{Specificity } = \#\text{True Negatives} / (\#\text{True Negatives} + \#\text{False Positives})$} is hindered by the high number of positions that do not present a fault event, the results are coherent with the sparse nature of the problem itself. Furthermore, BSS-Lasso's precision\footnote{$\text{Precision } = \#\text{True Positives} / (\#\text{True Positives} + \#\text{False Positives})$} of 77.87\% indicates that a selection does generally represent a true fault. This is extremely important from a practical point of view, as scheduling an in-field unit to repair a non-existing fault can be costly. At the same time, although the sensitivity\footnote{$\text{Sensitivity } = \#\text{True Positives} / (\#\text{True Positives} + \#\text{False Negatives})$} measure is roughly 80\%, analysis of the events which were not identified shows a two-fold behavior: either the fault magnitude was low and the BSS-Lasso neglects its presence; or the magnitude of an associated reflective event was low enough so that it was not identified and the shift correction was not employed, thus ensuing an error of over 50 m. In the first case, which corresponds to the majority of errors, the small faults do not entirely compromise the link's operation capacity. In the second case, though representing a more serious defect in fault location, the operator would have knowledge of the presence of the fault but with reduced accuracy. % Simulation Results

	%------------------------------------------------

	\section{Conclusions} \label{sec:conclusion}
	
	Manipulation of the detected backscattered Rayleigh signal inside the bandwidth of a frequency swept optical sub-carrier propagating into an optical fiber permits an efficient localization of faults through a Fourier operator. When the bandwidth is restricted, analysis in the frequency domain can overcome the spatial resolution limitation while also inducing a high-dimensional problem. Introducing the Lasso as a signal processing technique paired with the BSS framework, a methodology to consistently evaluate fiber defects, named BSS-Lasso, is designed.
    
	The BSS-Lasso presents several desired properties in the context of optical fiber link characterization, as corroborated by the results: (i) high specificity and sensitivity for reflective and non-reflective events; (ii) successful detection of multiple faults within a low-error interval; and (iii) low computational burden, having required less than two minutes for all links simulated and experimentally acquired in this work. Both experimental and simulation results show the efficacy of the \emph{ex-post} analysis, or correction stage, which besides accurately identifying the reflective events, also makes use of their presence to increase the precision of the fault locations. Furthermore, the BSS-Lasso framework allows for the reconstruction of the time-domain profile of fibers and the estimation of the magnitude of faults. In conclusion, the BSS-Lasso allows for precise, low-cost, transmitter-embedded full characterization of optical fiber links in practical computational time.
    
    We highlight several ideas that can be explored in future works to improve the proposed technique. From a data-acquisition point of view: (i) the analysis of the impact of the optical carrier's bandwidth and polarization on the measured frequency-domain profile; (ii) the extension of the frequency sweep range and whether this allows for increased estimation accuracy; and (iii) the employment of a lower NEP detector with the goal of extending the limited dynamic range of the BSS-Lasso. In the signal processing part: (i) determining whether the dynamic range is limited by an impossibility of acquiring signals from distant portions of the fiber or by the sparse characteristic of the Lasso, which neglects signal contributions that are too small; and (ii) identifying a more consistent methodology for the cluster analysis step. Finally, the possibility of embedding the software into a micro-controlled unit running Julia allied with a dedicated low-cost complex frequency beat detector would allow the creation of an independent measurement device, such as a standard OTDR, and is also a viable future development of this work.
    
    \section*{Acknowledgements}
    
    The authors would like to thank Christiano Nascimento and Breno Perlingeiro for technical support. This study was financed in part by the Coordenação de Aperfeiçoamento de Pessoal de Nível Superior - Brasil (CAPES) - Finance Code 001.
	
	%bibliografia
	\bibliographystyle{IEEEtran}
	\bibliography{References}
	
\end{document}